\documentclass[10pt,twocolumn,showpacs,longbibliography,amsmath,amssymb,osajnl,floatfix,superscriptaddress]{revtex4-1}

\usepackage{graphicx}
\usepackage{dcolumn}
\usepackage{bm}
\usepackage{color}
\usepackage{txfonts}
\usepackage{microtype}
\usepackage[T1]{fontenc}
\usepackage{lmodern}
\usepackage{subfigure}

\begin{document}

\author{Kun Xue}  \thanks{These authors have contributed equally to this work.}	\affiliation{MOE Key Laboratory for Nonequilibrium Synthesis and Modulation of Condensed Matter, School of Science, Xi'an Jiaotong University, Xi'an 710049, China}

\author{Zhen-Ke Dou}  \thanks{These authors have contributed equally to this work.}	\affiliation{MOE Key Laboratory for Nonequilibrium Synthesis and Modulation of Condensed Matter, School of Science, Xi'an Jiaotong University, Xi'an 710049, China}

\author{Feng Wan}	\affiliation{MOE Key Laboratory for Nonequilibrium Synthesis and Modulation of Condensed Matter, School of Science, Xi'an Jiaotong University, Xi'an 710049, China}

\author{Tong-Pu Yu}	\affiliation{Department of Physics, National University of Defense Technology, Changsha 410073, China}

\author{Wei-Min Wang}	\affiliation{Department of Physics
and Beijing Key Laboratory of Opto-electronic Functional Materials
and Micro-nano Devices, Renmin University of China, Beijing 100872,
China}
	
\author{Jie-Ru Ren}	\affiliation{MOE Key Laboratory for Nonequilibrium Synthesis and Modulation of Condensed Matter, School of Science, Xi'an Jiaotong University, Xi'an 710049, China}	

\author{Qian Zhao}	\affiliation{MOE Key Laboratory for Nonequilibrium Synthesis and Modulation of Condensed Matter, School of Science, Xi'an Jiaotong University, Xi'an 710049, China}	

\author{Yong-Tao Zhao}	\affiliation{MOE Key Laboratory for Nonequilibrium Synthesis and Modulation of Condensed Matter, School of Science, Xi'an Jiaotong University, Xi'an 710049, China}

\author{Zhong-Feng Xu}	\affiliation{MOE Key Laboratory for Nonequilibrium Synthesis and Modulation of Condensed Matter, School of Science, Xi'an Jiaotong University, Xi'an 710049, China}
	
\author{Jian-Xing Li}\email{jianxing@xjtu.edu.cn}
\affiliation{MOE Key Laboratory for Nonequilibrium Synthesis and Modulation of Condensed Matter, School of Science, Xi'an Jiaotong University, Xi'an 710049, China}

\title{Generation of highly-polarized high-energy brilliant $\gamma$-rays  via laser-plasma interaction}

\date{\today}

\begin{abstract}

Generation of highly-polarized high-energy brilliant $\gamma$-rays via laser-plasma interaction has been investigated in the quantum radiation-reaction regime. We employ a quantum-electrodynamics  particle-in-cell code to describe spin-resolved electron dynamics semiclassically and photon emission and polarization quantum mechanically in the local constant field approximation. As an ultrastrong linearly-polarized (LP) laser pulse irradiates on a near-critical-density (NCD)  plasma  followed by an ultrathin planar aluminum target, the electrons in NCD plasma are first accelerated by the driving laser to ultrarelativistic energies, and then  head-on collide with reflected laser pulse by the aluminum target, emitting brilliant LP $\gamma$-rays due to nonlinear Compton scattering with an average polarization of about 70\% and energy up to hundreds of MeV. By comparison, as a conical gold target filled with NCD plasma is employed, the linear polarization degree, collimation and brilliance  of emitted $\gamma$-ray beam are all significantly improved due to the enhanced strong laser-driven  quasi-static magnetic field in plasmas. Such $\gamma$-rays can be produced with currently achievable laser facilities and find various applications in high-energy physics and astrophysics.

\end{abstract}

\maketitle

\section{introduction}

Polarized high-energy $\gamma$-rays have a plenty of significant applications, e.g., generating polarized positrons and electrons \cite{Uggerhoj2005, Moortgat2008},  probing radiation mechanisms and properties of dark matter \cite{Bohm_2017} and black hole \cite{Laurent2011}, exciting polarization-dependent photofission of the nucleus in the giant dipole resonance \cite{Speth1981}, yielding meson-photoproduction \cite{Akbar2017}, and detecting vacuum birefringence in ultrastrong laser fields \cite{King_2016,Ilderton_2016,Ataman_2017,Nakamiya_2017,
Bragin2017}. 
Such $\gamma$-rays are commonly produced through 
either bremsstrahlung  \cite{Maximon_1959,Abbott_2016} or linear Compton scattering \cite{Omori_2006,Alexander_2008,Petrillo_2015}.
 However, the former can not generate linearly polarized (LP) ones and has deficiencies of large  scattering angle and emission divergence in the incoherent regime \cite{Baier1998}, and is limited by low current density of the impinging electrons and radiation flux due to the damage threshold of the crystal materials in the coherent regime~\cite{Lohmann1994,Carrigan_1987,Biryukov_1997}. While, the latter is severely restricted by the low electron-photon collision luminosity  due to the low laser intensities.

Nowadays, with rapid developments of strong laser techniques, state-of-the-art laser facilities can provide laser beams  with a peak intensity of about $10^{22}$ W/cm$^2$, pulse duration of about tens of femtoseconds and energy fluctuation $\sim$ 1\%
 \cite{Yoon2019,Danson_2019,Gales_2018,ELI,Vulcan,
 Exawatt,CORELS}, which stimulate experimental investigation on quantum electrodynamics (QED) processes during laser-plasma or laser-electron beam interactions \cite{Piazza2012, Blackburn_2019}. Such strong laser fields can be employed to directly polarize electrons \cite{Sorbo_2017,Sorbo_2018,Seipt_2018,
 Seipt_2019,li2019prl,Song_2019,Li_2019spin}
 due to radiative spin effects and create polarized positrons \cite{Chen_2019,Wan_2019plb} because of asymmetric spin-resolved pair production probabilities. Moreover, 
 in such strong laser fields Compton scattering process moves into the nonlinear regime: during the laser-electron interaction the electron radiates a high-energy $\gamma$ photon via absorbing millions of laser photons \cite{Piazza2012}. And, highly-polarized high-energy brilliant $\gamma$-rays can be generated via nonlinear Compton scattering in laser-electron beam interaction \cite{Ivanov_2004, King_2013, Ligammaray_2019}. In ultrastrong laser fields, the radiation formation length is much smaller than the laser wavelength and can not carry the driving laser helicity. Thus, the circular polarization of emitted $\gamma$-photons is transferred from the angular momentum  (helicity) of electrons  \cite{Ligammaray_2019}.  While, generating LP $\gamma$-photons does not require the electron polarization \cite{Wan_2020}.  Generated polarized high-energy $\gamma$-photons further interacting with the laser fields could produce electron-positron pairs
via multiphoton Breit-Wheeler process \cite{Breit_1934, Reiss1962,  Ritus_1985, Bula_1996}, 
and the photon polarization significantly affects the pair production probabilities \cite{Wan_2020, Breit_1934,Brown_1964, Nikishov_1964, Motz_1969, Baier1998, Kotkin_2005}.

Recently, all-optical $\gamma$-photon sources  have attracted broad interests \cite{Phuoc_2012, Chen_2013, Sarri2014, Yan_2017}. Usually,  brilliant high-energy $\gamma$-rays are generated experimentally or proposed theoretically  via bremsstrahlung \cite{Glinec_2005, Giulietti_2008}, nonlinear Thomson scattering \cite{Sarri2014, Yan_2017}, synchrotron \cite{Stark_2016, Benedetti_2018}, betatron \cite{Chang_2017, Yu_2013}, electron wiggling \cite{Wang_2018}, and nonlinear Compton scattering \cite{Zhu_2015,Zhu_2018, Gu_2018, Gu_2019}.
However, in those innovative works the information of the $\gamma$-photon polarization is overlooked hastily, which actually plays a significant role in the following secondary particles generation \cite{Wan_2020, Breit_1934,Brown_1964, Nikishov_1964, Motz_1969, Baier1998, Kotkin_2005}. Therefore, the polarization process of emitted $\gamma$-photons during laser-plasma interaction is still an open question.

In this paper, 
highly-polarized high-energy  brilliant $\gamma$-rays generated via laser-plasma interaction are studied in the quantum radiation-reaction regime with currently achievable laser intensities  \cite{Yoon2019,Danson_2019,Gales_2018,ELI,Vulcan,
 Exawatt,CORELS}. We implement the electron spin and photon polarization algorithms into the 
two-dimensional (2D) particle-in-cell (PIC)
EPOCH code \cite{Ridgers_2012, Arber_2015} in the local constant field approximation \cite{li2019prl, Ligammaray_2019, Chen_2019,Wan_2019plb, Wan_2020} to describe spin-resolved electron dynamics semiclassically and photon emission and polarization quantum mechanically. 
We first consider a commonly-used experimental setup: an ultrastrong  LP laser pulse irradiating on a near-critical-density (NCD) hydrogen plasma followed by an ultrathin planar aluminum (Al) target; see the interaction scenario in Fig.~\ref{fig1}.
The electrons in plasma are first  accelerated by the driving laser pulse  to ultrarelativistic energies, and then head-on collide with  reflected laser pulse  by the  Al target,
emitting abundant LP $\gamma$-photons via nonlinear Compton scattering with an average polarization  of about 70\%, energy up to hundreds of MeV and brilliance of the scale of $10^{21}$ photons/(s mm$^2$ mrad$^2$ 0.1\% BW) for the given parameters. Moreover, as a  conical gold (Au) target filled with NCD hydrogen plasma is employed instead of the NCD plus planar one, the polarization degree, collimation and brilliance of emitted $\gamma$-rays are all significantly improved because of the impact of  laser-driven strong quasi-static magnetic field; see the interaction scenario in Fig.~\ref{fig4}(a). Compared with the laser-electron beam interaction \cite{Ligammaray_2019}, this scenario is more accessible, since only a single laser beam is required. Besides, we also show the impact of the laser and target parameters on the $\gamma$-ray polarization and brilliance.

\section{simulation method}

In this work, we consider  the laser-plasma interaction in the quantum radiation-reaction regime, which requires a large  nonlinear QED parameter  $\chi_e\equiv |e|\sqrt{-(F_{\mu\nu}p^{\nu})^2}/m^3\gtrsim 1$ \cite{Ritus_1985, Koga2005}.
And, the multiphoton Breit-Wheeler pair production is characterized by another nonlinear QED parameter $\chi_{\gamma}\equiv |e|\sqrt{-(F_{\mu\nu}k_{\gamma}^{\nu})^2}/m^3$ \cite{Ritus_1985, Baier1998}. Here, $p$ and $k_\gamma$ are the 4-momenta of  electron and photon, respectively, $e$ and $m$ the electron charge and mass, respectively, and $F_{\mu\nu}$ the field tensor. 
As the electron head-on collides with the laser,  one can estimate $\chi_e\approx 2a_0\gamma_e\omega_0/m$, with the electron Lorentz factor $\gamma_e$ and the invariant laser field parameter
$a_0=eE_0/m\omega_0$. 
Here, $E_0$ and $\omega_0$ are the laser field amplitude and frequency, respectively.
Relativistic units with $c=\hbar=1$ are used throughout.

We implement the electron spin and photon polarization processes \cite{li2019prl, Ligammaray_2019, Chen_2019,Wan_2019plb, Wan_2020} into the 
2D-PIC
EPOCH code \cite{Ridgers_2012, Arber_2015}, in which we treat spin-resolved electron dynamics semiclassically, photon emission and pair production quantum mechanically in the local constant field approximation \cite{Ritus_1985,Baier1998,Ilderton2019prd, piazza2019},
which is valid at $a_0\gg 1$. At each simulation step, the photon emission and polarization are both electron-spin-dependent and calculated by following the Monte Carlo algorithms \cite{li2019prl, Ligammaray_2019, Wan_2020}, derived in the leading order contribution with respect to $1/\gamma_e$ via the QED operator method of Baier-Katkov \cite{Baier_1973}. Meanwhile, the photon polarization is represented by the Stokes parameters  ($\xi_1$, $\xi_2$, $\xi_3$), defined with respect to the axes $\hat{{\bf e}}_1=\hat{\bf a}-\hat{\bf v}(\hat{\bf v}\hat{\bf a})$ and $\hat{{\bf e}}_2=\hat{\bf v}\times\hat{\bf a}$ \cite{McMaster_1961}, with the photon emission direction $\hat{\bf n}$ along the electron velocity ${\bf v}$ for the ultrarelativistic electron (the emission angle $\sim1/\gamma_e\ll1$), $\hat{\bf v}={\bf v}/|{\bf v}|$,
and the unit vector $\hat{{\bf a}}={\bf a}/|{\bf a}|$ along the electron acceleration  ${\bf a}$. 
On detecting the mean polarization of a $\gamma$-photon beam, one must first normalize the Stokes
parameters of each photon to the same observation frame ($\hat{\textbf{o}}_1$, $\hat{\textbf{o}}_2$, $\hat{\textbf{n}}$),
i.e., rotate the Stokes parameters of each photon from its instantaneous frame ($\hat{\textbf{e}}_1$, $\hat{\textbf{e}}_2$, $\hat{\textbf{n}}$) to the same observation frame ($\hat{\textbf{o}}_1$, $\hat{\textbf{o}}_2$, $\hat{\textbf{n}}$), and then, calculate the average Stokes parameters of the $\gamma$-photon beam \cite{Ligammaray_2019, Wan_2020, McMaster_1961}.

After the photon emission the electron spin state is determined by the spin-resolved emission probabilities
and  instantaneously  collapsed into one of its basis states defined with respect to the instantaneous spin quantization axis (SQA), which is chosen according to the particular observable of interest: to determine the polarization of the electron along the magnetic field in its rest frame, the SQA is chosen along  the magnetic field ${\bm n}_B=\hat{\bf v}\times\hat{{\bf a}}$ 
\cite{li2019prl,Chen_2019}; in the case when the electron beam is initially  polarized with the initial spin vector ${\bf S}_i$, the observable of interest is the spin expectation value along the initial polarization and the SQA is chosen along that direction \cite{Ligammaray_2019}.
 Between photon emissions, the spin precession is governed by the Thomas-Bargmann-Michel-Telegdi  equation \cite{Thomas_1926, Thomas_1927, Bargmann_1959}. One can, alternatively, use the Cain code \cite{CAIN} to obtain uniform results.
 
As emitted high-energy $\gamma$-photons further interact with the strong laser fields, electron-positron pairs could be produced due to multiphoton Breit-Wheeler  process \cite{Breit_1934, Reiss1962,  Ritus_1985, Bula_1996}, and the pair production probabilities depend on the photon polarization \cite{Breit_1934,Brown_1964, Nikishov_1964, Motz_1969, Baier1998, Kotkin_2005}, which can be calculated by following the Monte Carlo method \cite{Wan_2020}.

\section{Polarization of $\gamma$-rays generated via laser-plasma interaction}

\subsection{Planar target}

\begin{figure}[t]
	\setlength{\abovecaptionskip}{-0.0cm} 	
\includegraphics[width=1\linewidth]{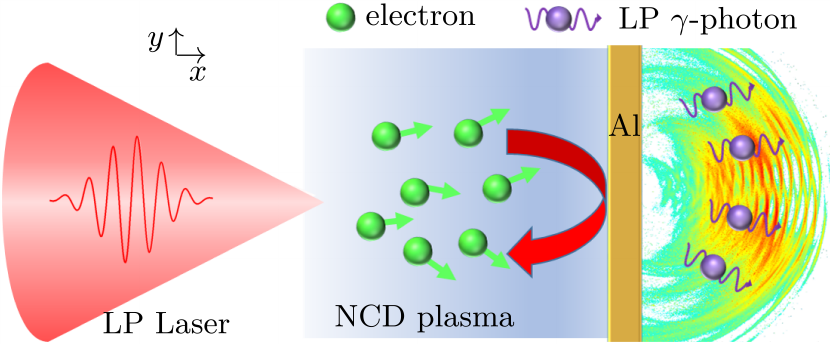}
	\caption{
	Scenario of generating LP $\gamma$-rays via nonlinear Compton scattering. An ultrastrong LP laser pulse, polarizing along $y$ axis and propagating along $x$ direction, irradiates on a NCD hydrogen plasma followed by an ultrathin planar Al target. Laser-driven ultrarelativistic electrons in plasma head-on collide with  reflected laser pulse by the Al target, emitting LP high-energy $\gamma$-photons, which can penetrate through the Al target and propagate forwards.
	}
	\label{fig1}
\end{figure}

We first consider a commonly-used experimental setup: an ultrastrong LP laser pulse irradiates on a NCD hydrogen plasma followed by an ultrathin planar Al target; see the interaction scenario in Fig.~\ref{fig1}. Electrons in plasma are accelerated to ulrarelativistic energies, and then head-on collide with reflected laser pulse by the Al target, emitting abundant LP high-energy $\gamma$-photons. 
To maximize the reflection of the driving laser pulse, the thickness of the Al target should be larger than the laser piston depth \cite{Robinson_2009}. Actually, as emitted $\gamma$-photons penetrate through the Al target, a proper thickness can also mitigate the bremsstrahlung and Bethe-Heitler pair production, which usually requires the thickness in the order of mm \cite{Erber_1966,Pike_2014}.

 Employed simulation box is $x\times y$ = $40\lambda_0 \times 30\lambda_0$, and  corresponding cells are $1000 \times 750$. The LP laser pulse injected from the left boundary polarizes along $y$ axis and propagates along $x$ direction with  wavelength $\lambda_0=1 \mu$m and normalized intensity $a = a_0$exp[-$(t-t_0)^2/\tau^2$]exp(-$y^2/w_0^2$), where focal radius $w_0=5\lambda_0$, pulse duration $\tau = 9T_0$  with laser period $T_0$,  corresponding full-width-at-half-maximum (FWHM) $\tau'=2\sqrt{\textrm{ln}2}\tau\approx$ 15$T_0$, and employed time delay $t_0=\tau$.  A solid planar Al target with electron density $n_e^{Al} = 702n_c$ and thickness $d_{Al}=1 \mu$m is placed at $20 \mu$m from the right boundary, where the plasma critical density $n_c = m\omega_0^2/4\pi e^2\approx 1.1\times10^{21}$ cm$^{-3}$. The left side of the Al target is filled with  NCD hydrogen plasma with electron density $n_e = 5n_c$ and thickness $d_p=10 \mu$m.  The numbers of macro-particles in each cell are 100 for electrons and 20 for H$^+$ and Al$^{3+}$ (fully ionized), respectively.

\begin{figure}[t]
	
	\setlength{\abovecaptionskip}{-0cm}
	\includegraphics[width=1.0\linewidth]{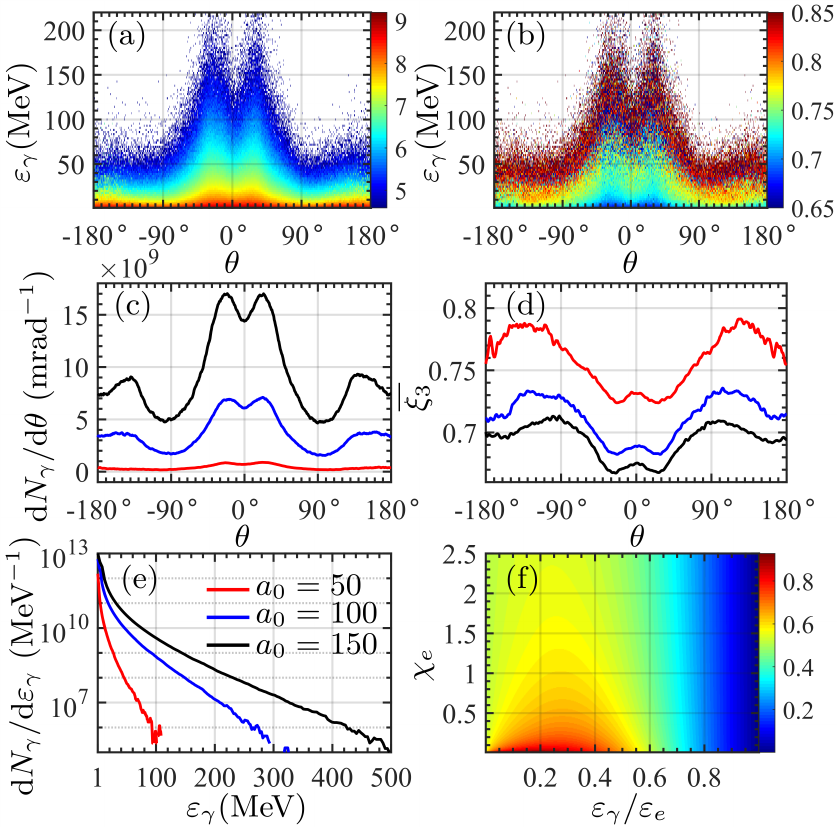}
	\caption{
		 (a) Angle-resolved density of emitted $\gamma$-rays
 log$_{10}$[d$^2N_\gamma$/(d$\varepsilon_\gamma$d$\theta$)] (MeV$^{-1}$ mrad$^{-1}$) vs the $\gamma$-photon energy $\varepsilon_{\gamma}$ and the polar angle $\theta$.
  (b) Linear polarization $\xi_3$ of emitted $\gamma$-photons vs $\varepsilon_{\gamma}$ and $\theta$. In (a) and (b), $a_0=100$. (c) and (d): 
  d$N_\gamma$/d$\theta$  and average linear polarization $\overline{\xi_3}$ vs $\theta$, respectively, calculated by summing over $\varepsilon_\gamma$ in (a) and (b), respectively. (e) d$N_\gamma$/d$\varepsilon_{\gamma}$  calculated by summing over $\theta$ in (a) vs  $\varepsilon_\gamma$. The red, blue and black curves in (c)-(e) indicate the cases of $a_0=$ 50, 100 and 150, respectively.  Only $\gamma$-photons with $\varepsilon_{\gamma} \ge 1$ MeV are counted. Other laser and target parameters are given in the text. (f) $\xi_3$ vs $\chi_e$ and $\varepsilon_\gamma/\varepsilon_e$.
	}
	\label{fig2}
\end{figure}

Distributions of the density and linear polarization of emitted $\gamma$-rays are illustrated in Figs.~\ref{fig2}(a) and (b), respectively, with the laser peak intensity $I_0\approx1.38\times 10^{22}$ W/cm$^2$ ($a_0=100$)  \cite{Yoon2019,Danson_2019,Gales_2018,ELI,Vulcan,
 Exawatt,CORELS}.  High-energy $\gamma$-photons of $\varepsilon_\gamma\gtrsim 100$ MeV are mainly emitted forwards, and two density peaks arise near $\theta\approx\pm 21^\circ$, since $\theta\propto a_R/\gamma_e\propto a_R/\varepsilon_\gamma$ (the electrons interact with reflected laser pulse with invariant intensity parameter $a_R$, and $a_R\propto a_0$), as shown in Fig.~\ref{fig2}(a), which is in excellent agreement with other simulations \cite{Lu_2020}. The linear polarization of $\gamma$-photons is characterized by the Stokes parameters $\xi_1$ and $\xi_3$ \cite{Ligammaray_2019, Wan_2020, McMaster_1961}. As we employ the basis vector of the observation frame $\hat{\bf o}_1$ in 
the polarization $y$-$x$ plane of the driving laser, according to the fact that $\gamma$-photons are mainly emitted in the polarization plane, $\xi_1$ is negligible and $\xi_3$ is given in Fig.~\ref{fig2}(b): at same polar angle the linear polarization $\xi_3$ for higher-energy $\gamma$-photons is relatively larger. The average linear polarization
$\overline{\xi_3}\approx0.68$ and partial $\xi_3$ can achieve up to 0.73.
In nonlinear Compton scattering the circular polarization defined by $\xi_2$ requires  initially longitudinally spin-polarized electrons \cite{Ligammaray_2019} and is negligible here.

To show the results in Figs.~\ref{fig2}(a) and (b) more visible, we sum over $\varepsilon_\gamma$ to obtain angle-resolved  number and linear polarization of emitted $\gamma$-photons; see the blue curves in Figs.~\ref{fig2}(c) and (d), and the energy density by summing over $\theta$ in Fig.~\ref{fig2}(a) is shown in Fig.~\ref{fig2}(e). 
In Figs.~\ref{fig2}(c)-(e) the impact of the driving laser intensity $a_0$  is studied. 
In ultrastrong laser fields, the number of the formation length in one laser period $T_0$ is  proportional to the laser intensity, and the photon emission probability in each formation length is  proportional to the fine structure constant $\alpha$ \cite{Piazza2012, Ritus_1985}, thus, the number of emitted $\gamma$-photons 
$N_\gamma\propto N_e\alpha a_R\tau/T_0\propto a_0$, where $N_e$ is the electron number. 
As $a_0$ increases from 50 to 150, $N_\gamma$ continuously rises up; see Fig.~\ref{fig2}(c). $\varepsilon_\gamma\propto \varepsilon_e \chi_e$, where the electron energy $\varepsilon_e\sim a_0$ (i.e., a stronger driving laser could accelerate the electrons to higher energies) and $\chi_e\propto a_R \gamma_e\propto a_0\varepsilon_e$,  thus, as $a_0$ increases, $\varepsilon_\gamma$ increases as well; see Fig.~\ref{fig2}(e). 

We underline that  $\overline{\xi_3}$ monotonically decreases with the increase of $a_0$, as shown in Fig.~\ref{fig2}(d). The physical reason is analyzed as follows. The Stokes parameter $\xi_3$ sensitively relies  on the parameters $\chi_e$ and $\varepsilon_\gamma/\varepsilon_e$ (see the analytical expression of $\xi_3$ in Refs.~\cite{Ligammaray_2019, Wan_2020}), as demonstrated  in Fig.~\ref{fig2}(f). 
As $\chi_e$  increases in the considered region, $\xi_3$ declines continuously; as $\varepsilon_\gamma/\varepsilon_e$ rises up, $\xi_3$ first  increases slightly and then gradually decreases to 0. Consequently, as $a_0\propto \chi_e$ increases, $\overline{\xi_3}$ declines in Fig.~\ref{fig2}(d), i.e., lower-intensity driving laser pulses can generate higher-polarization (but lower-brilliance) $\gamma$-photons.

For the case of $a_0=100$, the radius and duration of emitted $\gamma$-ray beam are $w_\gamma\approx w_0$ and $\tau_\gamma\approx\tau$, respectively. The angular divergences (FWHM) are approximately 
$1.74\times1.74$ rad$^2$, $1.47\times1.47$ rad$^2$, $1.39\times1.39$ rad$^2$ and $1.36\times1.36$ rad$^2$ for $\varepsilon_{\gamma}\ge 1$ MeV, 10 MeV, 100 MeV and 200 MeV, respectively. Corresponding brilliances are 
$0.67\times 10^{21}$, $2\times 10^{19}$, $7\times 10^{18}$ and $2\times 10^{17}$ photons/(s mm$^2$ mrad$^2$ 0.1\% BW) for $\varepsilon_{\gamma}= 1$ MeV, 10 MeV, 100 MeV and 200 MeV, respectively.
It is obvious that the brilliance $\propto N_r\propto a_0$. For the given parameters  multiphoton Breit-Wheeler pair production probabilities during $\gamma$-photons interacting with the laser fields are rather low, since $\chi_\gamma\ll1$.

\begin{figure}[t]
	\setlength{\abovecaptionskip}{-0cm} 		
	\includegraphics[width=1.0\linewidth]{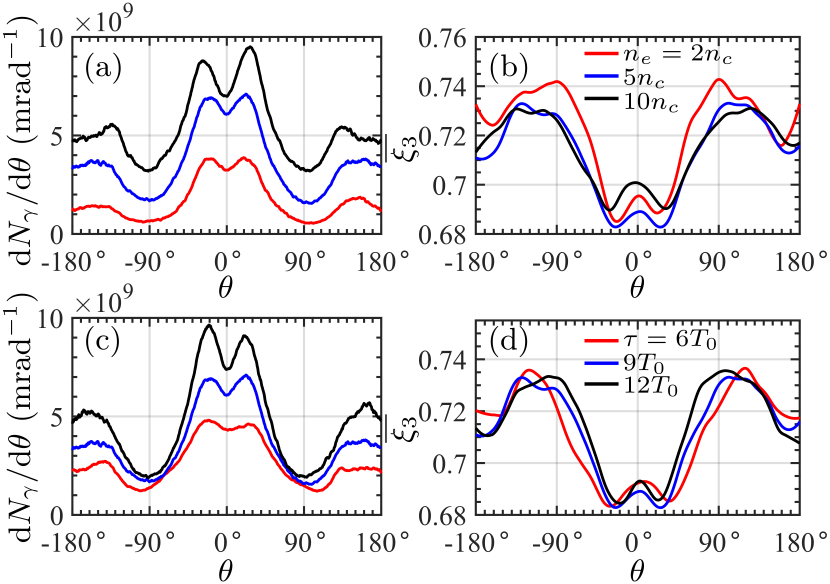}
	\caption{
(a) and (b): d$N_\gamma$/d$\theta$  and  $\overline{\xi_3}$ vs $\theta$, respectively, for the cases of $n_e=2n_c$ (red), $5n_c$ (blue) and $10n_c$ (black). (c) and (d): d$N_\gamma$/d$\theta$  and  $\overline{\xi_3}$ vs $\theta$, respectively, for the cases of $\tau=6T_0$ (red), $9T_0$ (blue) and $12T_0$ (black). $a_0=100$ and other laser and target parameters are the same with those in Fig.~\ref{fig2}.
	}\label{fig3}
\end{figure}

For the experimental feasibility, the impact of the laser and plasma parameters on the density and polarization degree  of emitted $\gamma$-photons is investigated in Fig.~\ref{fig3}. For instance, as the plasma density $n_e$ or the driving laser pulse duration $\tau$ increases, $N_\gamma\propto N_e\alpha a_R\tau/T_0\propto n_e \tau$ rises up accordingly; see Figs.~\ref{fig3}(a) and (c). However, 
for the NCD plasma, $\overline{\xi_3}$ changes slightly with the variations of $n_e$ and $\tau$; see Figs.~\ref{fig3}(b) and (d).
Note that as $n_e$ increases from $2n_c$ to $10n_c$, the opacity of the plasmas  increases as well and the laser propagation becomes more and more unstable, which results in an asymmetric angular distribution of emitted $\gamma$-rays \cite{Stark_2016}, as shown in Fig.~\ref{fig3}(a). Meanwhile, in plasma the driving laser pulse pushing the electrons in the low-density region forwards can yield a high-density region, where unstable laser propagation further induces  asymmetric  $\gamma$-photon emissions. This effect gets  severer as the laser pulse duration $\tau$ gets longer, as shown
 in Fig.~\ref{fig3}(c).
	
\subsection{Conical target}	
	
\begin{figure}[t]
		\setlength{\abovecaptionskip}{-0cm}
	\includegraphics[width=1.0\linewidth]{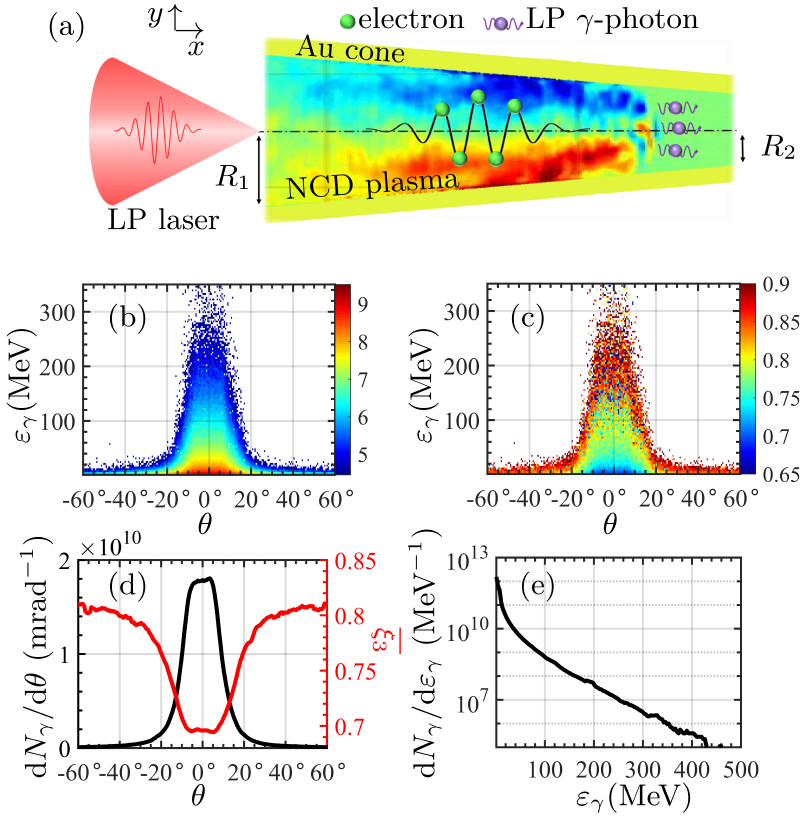}
	\caption{
		(a) Scenario of generating LP $\gamma$-rays via an ultrastrong LP laser pulse, polarizing along $y$ axis and propagating along $x$ direction, interacting with a conical Au  target filled with NCD hydrogen plasma. The red and blue areas inside the cone indicate the quasi-static magnetic fields along $+z$ and $-z$ direction, respectively, caused by the driving laser pulse. (b)
 log$_{10}$[d$^2N_\gamma$/(d$\varepsilon_\gamma$d$\theta$)] (MeV$^{-1}$ mrad$^{-1}$) vs  $\varepsilon_{\gamma}$ and  $\theta$.
  (c) $\xi_3$  vs $\varepsilon_{\gamma}$ and $\theta$.  (d) 
  d$N_\gamma$/d$\theta$ (black) and  $\overline{\xi_3}$ (red) vs $\theta$, respectively, calculated by summing over $\varepsilon_\gamma$ in (b) and (c), respectively. (e) d$N_\gamma$/d$\varepsilon_{\gamma}$  calculated by summing over $\theta$ in (b) vs  $\varepsilon_\gamma$.
 Only $\gamma$-photons with $\varepsilon_{\gamma} \ge 1$ MeV are counted. The results in (b)-(e) are at $t = 110T_0$, when the interaction has finished. The laser parameters are the same with those in Figs.~\ref{fig2}(a) and (b), and the target parameters are given in the text.
 }
	\label{fig4}
\end{figure}

\begin{figure}[t]
	\setlength{\abovecaptionskip}{-0cm}
	\includegraphics[width=1.0\linewidth]{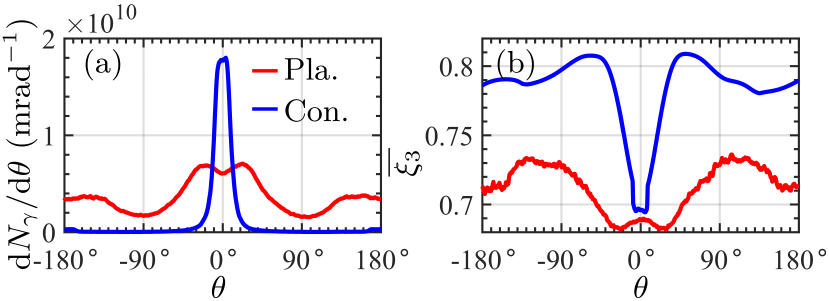}
\caption{ Comparisons of the results between the conical-target case (blue) in Fig.~\ref{fig4} and the planar-target case (red) in Fig.~\ref{fig2}, employing the same driving laser pulse. (a) and (b):  d$N_\gamma$/d$\theta$ and $\overline{\xi_3}$ vs $\theta$, respectively.
}
	\label{fig5}
\end{figure}

To improve the polarization, collimation and brilliance of emitted $\gamma$-rays, we employ a conical Au target filled with NCD hydrogen plasma, instead of the planar target in Figs.~\ref{fig1} and \ref{fig2}; see the interaction scenario in Fig.~\ref{fig4}(a). The front and rear surfaces of the cone are open. As an ultraintense LP laser pulse irradiates on the NCD hydrogen plasma inside of Au cone, almost all bulk electrons are pushed forwards and excite a strong quasi-static magnetic field $B_p$ \cite{Zhu_2015,Stark_2016}. The maximum intensity of the magnetic field can be estimated as Max(${B_p}$)$\simeq 4\pi|e|\beta_e n_e  R$, which is of the same order of the magnetic field of the driving laser \cite{Stark_2016}. $R$ and $\beta_e$ denote the radius of the cone and the electron velocity scaled by the light speed in vacuum. For the plane-wave case, the transverse electric field ${\bf E_\perp}$ can almost cancel ${\bf v\cdot B}$ with  the magnetic field ${\bf B}$, and consequently, $\chi_e$ is rather small and the photon emissions are very weak. While, in the conical-target case with a strong quasi-static magnetic field ${B_p}$, $\chi_e\propto B_p$  greatly increases, thus,  subsequent $\gamma$-photon emissions are significantly enhanced. Moreover, since most photons are emitted at the edge of the cone, where the tansverse velocities of electrons are close to 0, the angular spread of the $\gamma$-ray beam is narrowed \cite{Stark_2016}.

\begin{figure}[t]
	\setlength{\abovecaptionskip}{-0cm}
	\includegraphics[width=1.0\linewidth]{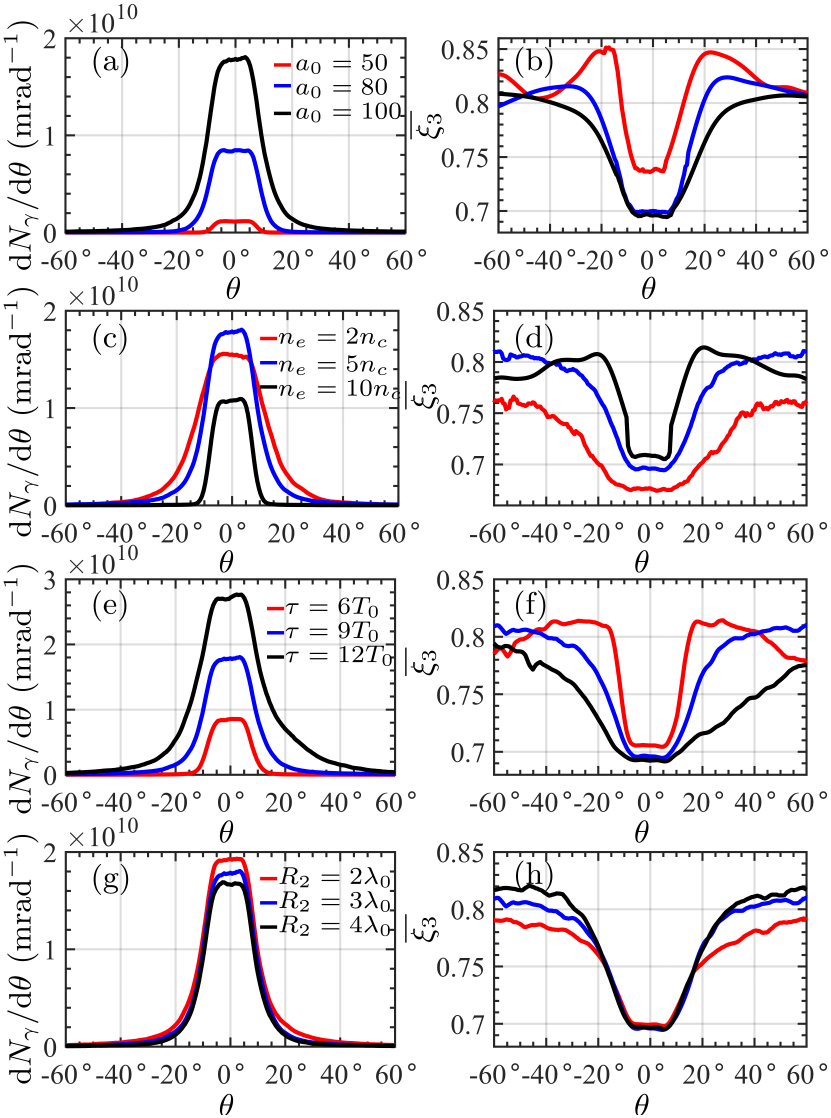}
	\caption{ 	
Left and right columns:  d$N_\gamma$/d$\theta$ and $\overline{\xi_3}$ vs $\theta$, respectively.
The red, blue and black curves indicate in (a) and (b) $a_0$ = 50, 80 and 100, respectively; in (c) and (d) $n_e$ = $2n_c$, $5n_c$ and $10n_c$, respectively; in (e) and (f) $\tau=6T_0$, $9T_0$ and $12T_0$, respectively; in (g) and (h) $R_2= 2\lambda_0$, $3\lambda_0$ and $4\lambda_0$, respectively. Other laser and target parameters are the same with those in Fig.~\ref{fig4}.
	}
	\label{fig6}
\end{figure}

	In Fig.~\ref{fig4} the simulation box is $x\times y$ = $110\lambda_0 \times 20\lambda_0$, and corresponding cells are $2750\times500$. A solid  conical Au target with electron density $n_e^{Au} = 100n_c$ and thickness $d_{Au}=1 \mu$m is located in the region from $5\lambda_0$ to $75\lambda_0$ in the $x$ axis. The left and right opening radii of the cone are $R_1 = 7 \lambda_0$ and $R_2 = 3 \lambda_0$, respectively. The Au cone is filled with NCD hydrogen plasma with  density  $n_e = 5n_c$. The numbers of macro-particles in each cell are 100 for electrons and 20 for H$^+$ and Au$^{2+}$ (partially ionized), respectively.

The density and linear polarization of emitted  $\gamma$-photons in the conical-target case are represented in Figs.~\ref{fig4}(b)-(e), and  comparisons of the  results between the conical-target case in Fig.~\ref{fig4} and the planar-target case in Fig.~\ref{fig2}, employing the same driving laser pulse, are illustrated in Fig.~\ref{fig5}. 
In the conical-target case, most photon emissions are induced by the quasi-static magnetic field at the edge of the cone, which is weaker than reflected laser field in the planar-target case. Thus, for the former $\chi_e$ is smaller, and consequently, $\xi_3$ is larger (see the analysis on the relationship of $\xi_3$ to $\chi_e$ in Fig.~\ref{fig2}(f)), as shown in Fig.~\ref{fig5}(b). $\overline{\xi_3}\approx0.71$ and partial $\xi_3$ can achieve up to about 0.81 for the given parameters. 
The angular divergences (FWHM) are approximately 
$0.35\times0.35$ rad$^2$, $0.31\times0.31$ rad$^2$, $0.28\times0.28$ rad$^2$, $0.24\times0.24$ rad$^2$ for $\varepsilon_{\gamma}\ge 1$ MeV, 10 MeV, 100 MeV and 200 MeV, respectively. The  corresponding brilliances are 
$1.29\times 10^{21}$, $1.10\times 10^{21}$, $0.92\times 10^{20}$ and $1.5\times 10^{19}$ photons/(s mm$^2$ mrad$^2$ 0.1\% BW) for $\varepsilon_{\gamma}= 1$ MeV, 10 MeV, 100 MeV and 200 MeV, respectively. Even though the target parameters are not exactly the same for these two cases,  the collimation and brilliance of emitted $\gamma$-photons are  both qualitatively improved in the conical-target case. 

The impact of the laser and target parameters on the density and polarization of emitted $\gamma$-rays is investigated comprehensively in Fig.~\ref{fig6}. As $a_0$ or $\tau$ increases, $N_\gamma\propto n_e B_p\tau\propto n_e a_0 \tau$ increases as well, as shown in Figs.~\ref{fig6}(a) and (e), however, $\overline{\xi_3}$ decreases due to the increase of $\chi_e\propto B_p \gamma_e\propto a_0 n_e  \tau$ ($\gamma_e\propto \tau$, i.e., the electrons obtain more energies in longer laser pulses), as shown in Figs.~\ref{fig6}(b) and (f).
The increase of $n_e$  can enhance the opacity of the plasma, and the interaction regime will transit from relativistically transparent to relativistically NCD. For instance,
as  $n_e$ rises up continuously from $2n_c$ to $10n_c$, the stochastic electron heating is enhanced (the charge separation field and the quasi-static magnetic field become stronger) and the electron acceleration is weakened ($\gamma_e$ declines), thus,
$\chi_e \propto \gamma_e n_e$  first rises up due to the increase of $n_e$ and then goes down because of the decrease of $\gamma_e$, which results in corresponding variations of $N_\gamma$ and $\overline{\xi_3}$  in Figs.~\ref{fig6}(c) and (d).
As the rear radius of the conical target $R_2$ rises up, the laser focusing effect is weakened, which induces  the decrease of the strength of the quasi-static magnetic field $B_p$ and $\chi_e\propto B_p$. Therefore, 
$N_\gamma\propto  B_p$ declines in Fig.~\ref{fig6}(g) and $\overline{\xi_3}$ increases in Fig.~\ref{fig6}(h).

\section{conclusion}
Generation of highly-polarized high-energy brilliant $\gamma$-rays is studied via employing a single-shot ultraintense LP laser pulse interacting with solid planar and conical targets filled with NCD plasma in the quantum radiation-reaction regime, and with currently achievable laser intensities  emitted $\gamma$-rays can reach  an average linear polarization of about 70\%, energy up to hundreds of MeV and brilliance of the scale of $10^{21}$ photons/(s mm$^2$ mrad$^2$ 0.1\% BW), 
which can be applicable in high-energy physics and astrophysics. Moreover, those considered interaction regimes are confirmed to be robust with respect to the laser and target parameters.\\

\textbf{ACKNOWLEDGMENTS}\\
This work is supported by the National Natural Science Foundation of China (Grants Nos. 11874295, 11875219, 11705141, 11905169 and  11875319), the National Key R\&D Program of China (Grant Nos. 2018YFA0404801 and 2018YFA0404802), Chinese Science Challenge Project No. TZ2016005, National Key Research and Development Project No. 2019YFA0404900

\bibliography{QEDspin}
\end{document}